\documentclass[10pt,prd,tightenlines,twocolumn,superscriptaddress]{revtex4}

\usepackage{amsmath,amssymb}

\usepackage[dvips]{graphicx,color}
\definecolor{darkblue}{rgb}{0,0,0.7}
\definecolor{darkred}{rgb}{0.7,0,0}
\usepackage[dvips,unicode, colorlinks, citecolor=darkblue, linkcolor=darkred,urlcolor=blue]{hyperref}

\usepackage{psfrag}

\begin{document}

\title{Stable double-resonance optical spring in laser gravitational-wave detectors}

\author{Andrey A. Rakhubovsky}
\affiliation{Physics Department, Moscow State University, Moscow 119992 Russia}

\author{Stefan Hild}
\affiliation{SUPA, School of Physics and Astronomy, University of Glasgow, Glasgow, Q12 8QQ, UK}

\author{Sergey P. Vyatchanin}
\affiliation{Physics Department, Moscow State University, Moscow 119992 Russia}
\date{\today}

\begin{abstract}
We analyze the optical spring characteristics of a double pumped Fabry-Perot
cavity. A double-resonance optical spring occurs when the optical spring frequency and the detuning
frequency of the cavity coincide. We formulate a simple criterion for the stability of an optical spring and apply it to the double resonance regime. Double resonance configurations are very promising for future gravitational wave detectors as they allow us to surpass the Standard Quantum Limit. We show that stable double resonance can be demonstrated in middle scale prototype interferometers such as the  Glasgow 10m-Prototype, Gingin High Optical Power Test Facility or the AEI 10m Prototype Interferometer before being implemented in future gravitational wave detectors.

\end{abstract}

\maketitle

\section{Introduction}

Currently the search for gravitational radiation from astrophysical
sources is conducted with the first-generation Earth-based laser
interferometers (LIGO in USA \cite{1992_LIGO,2006_LIGO_status,website_LIGO}, VIRGO in Italy
\cite{2006_VIRGO_status,website_VIRGO}, GEO-600 in Germany
\cite{2006_GEO-600_status,website_GEO-600}, TAMA-300 in Japan
\cite{2005_TAMA-300_status,website_TAMA-300} and ACIGA in Australia
\cite{2006_ACIGA_status,website_ACIGA}). The development of the
second-generation GW detectors (Advanced LIGO
\cite{2002_Adv_LIGO_config,website_Adv_LIGO}, Advanced Virgo
\cite{website_Advirgo}, GEO-HF \cite{GEO-HF} and LCGT
\cite{2006_LCGT_status}) is well underway.

The sensitivity of the first-generation detectors is limited by noises sources of various nature: seismic and suspension thermal noise
noise at low frequencies (below $\sim 50$ Hz), thermal noise in suspensions, bulks and coatings of the mirrors
 ($\sim 50-200$ Hz), photon shot noise (above $\sim 200$ Hz). It is expected that the sensitivity of the second-generation
 detectors will be ultimately limited by the noise of quantum nature arising due to Heisenberg's uncertainty principle
\cite{1968_SQL,1975_SQL,1977_SQL,1992_quant_meas} over most of the frequency range of interest. The optimum between
 measurement noise (photon shot noise) and back-action noise (radiation pressure noise) is called the Standard Quantum Limit (SQL).
This level is expected to be reached in the forthcoming second generation of large-scale laser-interferometric gravitational-wave detectors.
Third generation detectors, such as the Einstein Telescope \cite{et_punturo2010} aim to significantly surpass the SQL over a wide
 frequency range \cite{Hild10}.

\begin{figure}[t]

\psfrag{xE}{$x_E$}
\psfrag{yE}{$y_E$}
\psfrag{xN}{$x_N$}
\psfrag{yN}{$y_N$}

\psfrag{1}{$L_1$}
\psfrag{2}{$L_2$}

\psfrag{aP}{}
\psfrag{bP}{}
\psfrag{m}{$m$} \psfrag{rho}{$\rho$}
\psfrag{mT}{$m,\, T$}

\psfrag{ad1}{} \psfrag{ad2}{}
\psfrag{bd1}{} \psfrag{bd2}{}
\psfrag{aS}[lc][lb]{$a_s$}
\psfrag{bS}[rc][lb]{$b_s$}

\psfrag{aE}[cb][lb]{$a_E$}
\psfrag{bE}[ct][lb]{$b_E$}
\psfrag{aE1}[lt][rb]{$a_{E1}$}

\psfrag{bE1}[lb][rb]{$b_{E1}$}
\psfrag{aE2}[cb][lb]{$a_{E2}$}
\psfrag{bE2}[ct][lb]{$b_{E2}$}

\psfrag{aN}[rc][lb]{$a_N$}
\psfrag{bN}[lc][lb]{$b_E$}
\psfrag{aN1}[lc][lt]{$a_{N1}$}
\psfrag{bN1}[rc][lb]{$b_{N1}$}
\psfrag{aN2}[rc][lb]{$a_{N2}$}
\psfrag{bN2}[cc][cc]{$b_{N2}$}
\psfrag{SRM}{SRM}
\psfrag{PRM}{PRM}

\includegraphics[width=0.45\textwidth,height=0.4\textwidth]{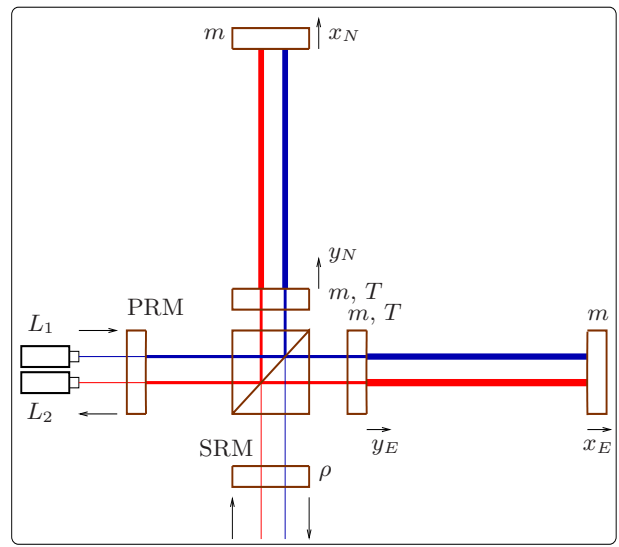}
\caption{Scheme of an Advanced LIGO interferometer pumped by two lasers. The main laser is detuned to give a positive
 optical rigidity and negative damping (it is tuned on the right slope of resonance curve)  while the auxiliary  laser is detuned
to give negative optical rigidity and positive  damping.
}
\label{fig:AdvLIGO}
\end{figure}

The most promising methods to overcome the SQL rely on the implementation of optical (ponderomotive)
rigidity \cite{64a1BrMiVMU,67a1BrMaJETP,78BrBook, 1992_quant_meas}
which effectively turns the test masses of a  gravitational-wave detector into
harmonic oscillators  producing a gain in sensitivity
\cite{99a1BrKhPLA,01a1KhPLA,01ChenPRD,02ChenPRD,05a1LaVyPLA,06a1KhLaVyPRD}.
In order to understand it note that the formula for the sensitivity $\xi(\Omega)$ of Advanced LIGO
interferometer (see Fig.~\ref{fig:AdvLIGO})   consists of two terms:
\begin{align}
\label{xisq}
\xi^2(\Omega) &= \frac{S_h(\Omega)}{h_{\rm SQL}^2(\Omega)}=\\
&=	\frac{2}{\hbar m\Omega^2}\left(S_F(\Omega) +
        	\mu^2\big|K(\Omega)-\Omega^2\big|^{2}S_x(\Omega)\right).\nonumber
\end{align}
Here $S_h$ is the total single-sided spectral density of the noise recalculated to the strain density $h$ of the gravitational wave
$h_{\rm SQL}^2= 2\hbar/\mu L^2\Omega^2$ --- the value of  $S_h$ corresponds to the SQL sensitivity for the case of
free masses, $\Omega$ is the observation frequency, $L$ the arm length of the interferometer, $\mu=m/4$ is the reduced
mass, $m$ is the mass
of each mirror of the arm cavities, $S_x$ and $S_F$ are the single-sided spectral densities of the measurement noise and the back
action noise, respectively and $\mu K(\Omega)$ is the optical rigidity. The relation between the
 Fourier transforms of the position $x$ and the force $F$ can be described as
$\mu\big(-\Omega^2+K(\Omega)\big)x(\Omega)=F(\Omega)$. For  simplicity we assume here
that the measurement and the back action noise are not correlated. Then the uncertainty relation
is given by
\begin{align}
\label{SfSx}
 S_F(\Omega)\, S_x(\Omega)& \ge \hbar^2\,.
 \end{align}
Substituting (\ref{SfSx}) in form S$_F(\Omega)\ge \hbar^2/\, S_x(\Omega)$ into (\ref{xisq}) we can find the
minimal value of the sensitivity $\xi(\Omega)$ (after optimization over $S_x$) which is better than the SQL
\begin{align}
\xi^2_\text{min}(\Omega) &= \frac{|-\Omega^2+K(\Omega)|}{\Omega^2}< 1
\end{align}
in the bandwidth $\Delta \Omega$  where $|K(\Omega)-\Omega^2|< \Omega^2$, i.e. close to
optical spring resonance.
In  other words a harmonic oscillator provides a gain in sensitivity for near-resonance signals, equal to
\begin{equation}
\label{xi_oscill}
  \xi^2 = \frac{\Delta\Omega}{\Omega} < 1 \,.
\end{equation}
Usually the two resonances are at two separate frequencies.
The gain in sensitivity by means of the optical rigidity was examined in
 \cite{01ChenPRD,02ChenPRD,04ChenPRD} for the wide-band regime under
conditions $\Delta\Omega\sim\Omega$ and $\xi\gtrsim 1/2$.

The case of so-called \emph{double resonance} \cite{01a1KhPLA, 05a1LaVyPLA, 06a1KhLaVyPRD}, when
an sophisticated frequency dependence of the optical rigidity allows us to obtain two
close or coinciding resonance frequencies and to get better gain in sensitivity, is described by the following
formula:
\begin{equation}
\label{xi_fdrigid}
  \xi^2 = \frac{S_{h}}{h_{\rm SQL}^2}
    = \left(\frac{\Delta\Omega}{\Omega}\right)^2 \,.
\end{equation}
It is important to note that the oscillatory sensitivity gain, described in Equation (\ref{xi_oscill}) does not provide
any gain in the signal to noise ratio in case the signal has a bandwidth larger than $\Delta\Omega$
(because the signal to noise ratio scales as the sensitivity gain multiplied by
the bandwidth $\Delta\Omega$). However, for a double resonance configuration with a sensitivity described by Equation~(\ref{xi_fdrigid}) the signal
to noise ratio increases with decreasing bandwidth proportional to $\Delta\Omega^{-1}$ \cite{01a1KhPLA, 05a1LaVyPLA, 06a1KhLaVyPRD}.
Please note  that in the Advanced LIGO configuration  optical rigidity can be created  easily through microscopic position
 changes of the signal recycling mirror \cite{01ChenPRD,02ChenPRD,05a1LaVyPLA, 06a1KhLaVyPRD} as it is already being
used in GEO\,600 detector \cite{2006_GEO-600_status}.

A single optical spring always causes instability. This can qualitatively
be explained \cite{64a1BrMiVMU,67a1BrMaJETP,78BrBook} by taking into account
that the optical rigidity is not introduced instantaneously, but with a delay of the
 relaxation time $\tau^*$ of the optical resonator. Therefore, the evolution of the free mass
position $x$ with the
 optical rigidity $K$ may approximately be	described by the equation
\begin{equation}
\label{xEvol}
m\ddot x(t) +K x(t-\tau^*)=0.
\end{equation}
Expanding the second term in a series $ x(t-\tau^*)\simeq x(t) -\tau^*\dot x(t)$ we can
rewrite the previous equation in the form:
\begin{equation}
\label{xEvol2}
m\,\ddot x(t) -K\tau^*\, \dot x(t) +K\,x(t)=0.
\end{equation}
Obviously, the term $-K\tau^*\, \dot x$ corresponds to a negative damping force. Please note that in
the opposite case of {\em negative} optical rigidity a {\em positive} damping force is introduced
(the sign of the rigidity correlates with the sign of the detuning).

The instability (negative damping) can be compensated by incorporating a linear feedback
control loop and in the ideal case (no additional noise is introduced by the feedback) it would not
modify the noise spectrum of a GW detector \cite{02ChenPRD}.  In practice, however, the need for control
gain at frequencies inside the detection band can cause undesirable complexity in the control system or can introduce additional
classical noise.

An alternative way to suppress the instability was proposed \cite{08ChenPRD} and experimentally demonstrated
\cite{07CorbitPRL}, by injecting a second carrier field from the bright port (see Fig.~\ref{fig:AdvLIGO}) in order
 to create a relatively small additional negative rigidity component, thus the total rigidity (of both lasers together) remains
 positive, but at the same time introduce a relatively large additional positive damping component to make total damping positive.
 The main purpose of the second carrier is to create a second optical spring that forms a stable optical spring together with the first
 one --- even though each individual optical spring, acting alone, would be unstable. Both carriers are assumed to have different
 polarizations, so that there is no direct coupling between the two fields (although they both directly couple to the mirrors).

In this paper we further analyze the stability of an optical spring created by a double pump. The regime of {\em stable} double
resonance analyzed in this paper provides the possibility to decrease the bandwidth of the resonance curve dramatically
and, hence, to increase the signal to noise ratio even for signals with a  wide bandwidth (as formula (\ref{xi_fdrigid}) predicts).
In Sec.~\ref{SOP} we formulate the simple criterion of stability and we  apply it to the regimes of stable double resonance in
Sec.~\ref{SDR}.
In addition we discuss the possibilities to observe these regimes using the experimental setups of the Gingin and Glasgow Prototype
interferometers.

\section{Stable optical spring}\label{SOP}
In the following we use notations similar to the ones introduced in \cite{01ChenPRD, 02a1KiLeMaThVyPRD}.
We  consider an Advanced  LIGO interferometer with a signal recycling
mirror (SRM) having an amplitude reflectivity $\rho$ and a power recycling  mirror (PRM)
as shown in Fig.\ref{fig:AdvLIGO}. We assume the mirrors to be without any optical losses.
In addition, we suppose that both Fabry-Perot (FP) cavities in the east and north arms are
identical, each  input mirror featuring an amplitude transmittance $T\ll 1$ and an reflectivity
$R=\sqrt{1-T^2}$, while each end mirror is completely reflecting. The  end
and input mirrors have an identical mass $m$ and in absence of the laser pumps they can move as free masses.
 The PRM is used only to increase the average power incident onto the beam splitter
and we assume that no fluctuational fields from the laser (west arm) reach the detector in the south
arm. The mean frequency of each pump  $\omega_{1,2}$ is equal to one of the eigen-frequencies of the FP cavities.
 $L$ is the distance between the mirrors in the arms ($4$~km for (Advanced) LIGO), $l$ is the mean
distance between the SR mirror and the beam splitter. We also introduce the following notations:
\begin{align}
\label{gamma0}
 e^{i\Omega L/c}& \simeq 1+\frac{i\Omega L}{c},
 	\quad \gamma_0=\frac{cT^2}{4L},\\
\phi_{1,2} &= \frac{(\omega_{1,2}+\Omega) l}{c}\simeq\frac{\omega_{1,2} l}{c}, \\
\label{Gamma}
\Gamma_{1,2} &= \gamma_0\,\frac{1-\rho^2}{1+2\rho\cos2\phi_{1,2}+\rho^2}\,,\\
\label{Delta}
\Delta_{1,2} & = \gamma_0\,\frac{2\rho\sin 2\phi_{1,2}}{1+2\rho\cos2\phi_{1,2}+\rho^2}\,,\
\end{align}
Here $\gamma_0$ is the relaxation rate of a single FP cavity, $\Delta_{1,2}$ are the detunings
introduced by the SR mirror, $\Gamma_{1,2}$ are the relaxation rates of the  differential modes of the interferometer for each pump.
The frequency of the differential mode depends on the  arm length difference $z = x_E - y_E-(x_N - y_N)$ and its
 bandwidth and detuning are controlled by position of SR mirror, which is described by phase advance $\phi_{1,2}$
arising from differential mode detuning ($\phi_{1,2}=0$ corresponds to
resonance).
We assume that $\phi_{1,2}$ do not depend on $\Omega$ due to the small length $l$ of the SR cavity: $l \ll L$.
It is worthwhile noting that the detunings $\Delta_{1,2}$ and relaxation rates $\Gamma_{1,2}$ can differ for different pumps.

The mechanical evolution of the differential arm length
$z=x_E-y_E -x_N+y_N$ (see notations on Fig.~\ref{fig:AdvLIGO})  in the frequency domain is described by
\begin{subequations}
 \label{z}
\begin{align}
z =& \Psi(\Omega)\big(F_1+F_2 +F_s \big),\quad F_s= \mu\Omega^2 L\, h  ,\\
\label{mechI}
\Psi(\Omega) &= \frac{1}{\mu\big[-\Omega^2 + K_1(\Omega) + K_2(\Omega)\big]},\quad
	\mu=\frac{m}{4},\\
K_{1,2}&= 	\frac{ 8 I_{1,2}\, \omega_{1,2}\,  \Delta_{1,2}}{cmL\,
	\left[\big(\Gamma_{1,2} -i\Omega\big)^2+\Delta_{1,2}^2\right]} ,
\end{align}
\end{subequations}
where $h$ is a dimensionless metric of a gravitational wave, $\Psi$ is the mechanical susceptibility
($K_{1,2}$ are the frequency dependent spring coefficients
created by the corresponding laser pump); $I_{1,2}$ are the average optical arm cavity powers of the carrier
 $1$ or $2$ with the mean optical frequencies $\omega_{1,2}$;  $F_1,\ F_2$
are the back action forces caused by the first and second pump.

The eigen frequencies are solutions of the characteristic equation
\begin{align}
\label{Omega}
 \Psi(\Omega)^{-1}&=0
\end{align}
Roots corresponding to stable oscillations must have  a {\em negative} imaginary part (as we assume the time dependence of position to be $\sim e^{-i\Omega t}$).

The equation (\ref{Omega}) can be rewritten using the dimensionless susceptibility
$\chi=\Psi\, m(\Gamma_1^2+\Delta_1^2)/8$ (here $m(\Gamma_1^2+\Delta_1^2)/8$ is a convenient multilpier):
\begin{subequations}
\label{notations2}
\begin{align}
\chi^{-1} &= \frac{Y}{\big(2-x^2-ig x\big)\big(2\eta -x^2-i\nu g x\big)}\,, \\
\label{AligoY}
Y & =Y_r+ig x Y_i\,,\quad \\
\label{AligoYr}
Y_r &= x^6- \big[2(1+\eta )+\nu g^2\big]x^4 +\\
 &\quad +(4\eta +P+Q) x^2 -2\eta P-2Q\,,\nonumber\\
 \label{AligoYi}
Y_i &=\big(1+\nu\big) x^4 -2(\eta +\nu )\, x^2 +\nu P + Q\, ,\\
 x& \equiv  \frac{\sqrt 2\,\Omega}{ \sqrt{\Gamma_1^2+\Delta_1^2}},\quad
 g\equiv \frac{2\sqrt 2\, \Gamma_{1}}{\sqrt{\Gamma_1^2+\Delta_1^2}}\, ,\\
 P & \equiv  \frac{32\omega_1I_1\Delta_1}{m cL\big(\Gamma_1^2+\Delta_1^2\big)^2},\quad
 Q \equiv \frac{32 \omega_2I_2\Delta_2}{mc L \big(\Gamma_1^2+\Delta_1^2\big)^2},\nonumber\\
 \eta &\equiv \frac{\Delta_2^2+\Gamma_2^2}{\Delta_1^2+\Gamma_1^2},\quad
 	\nu\equiv \frac{\Gamma_2}{\Gamma_1},.
\end{align}
\end{subequations}
We choose the dimensionless frequency $x$ so that for a single pump (i.e. $Q=0$) the
double (unstable) resonance would take place at $x^2\simeq 1$ if  $P=1$ \cite{05a1LaVyPLA} (in formal limit  $g\to 0$ when susceptibility is a pure real value).

\begin{figure}[ht]
\psfrag{Yr}{$Y_r$} \psfrag{Yi}{$Y_i$} \psfrag{c}{\bf c)} \psfrag{d}{\bf d)}
\psfrag{x2}{$x^2$} \psfrag{x02}{$x_0^2$}
\psfrag{x01}{$\left[x_1^{(0)}\right]^2$}
\psfrag{x002}{$\left[x_2^{(0)}\right]^2$} \psfrag{x03}{$\left[x_3^{(0)}\right]^2$}
 \includegraphics[width=0.33\textwidth, height=0.25	\textwidth]{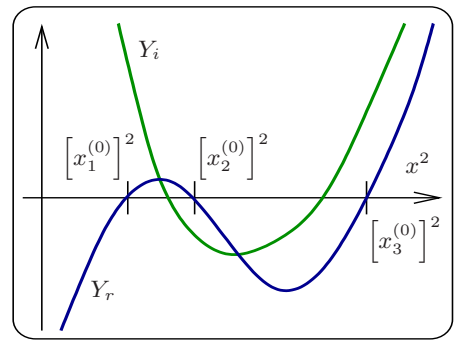}
 \caption{Roots correspond to a stable (slightly damped) oscillation if $Y_r'$ and $Y_i$
have the same signs near roots of $Y_r$, i.e.  plots of $Y_r$ and $Y_i$ correspond to
each other as shown on plot.}\label{stable}
\end{figure}

In order to find the eigen-frequencies one has to solve the equation $Y(x)=0$ and
to analyze for which conditions the roots are stable. Undoubtedly the roots of this equation can easily be
obtained by applying numerical methods. However, starting from a set of input parameters (powers and detunings
of each pump) the numerical solution will provide a set of frequencies with {\em arbitrary} imaginary parts.
The problem is to identify the values of the system parameters that actually provide stable roots
with {\em negative} imaginary parts.  In the following we formulate a criterion which allows us in a quite easy way to
properly estimate a set of  parameters fulfilling our requirements.

In the general case, $Y_r$ is a third order polynomial of $x^2$, so all of it's roots
(which we assume to be real) can be written in the form of $\pm x_j^{(0)}\ (j = 1,2,3).$
An example of $Y_r$ as a function of $x^2$ is plotted on Fig.~\ref{stable}.

In order to apply a successive approximation method we assume that the values of the function $Y_i$ are small
in the regions close to the roots of $Y_r$. If this is the case, then the roots of
the equation $Y(x) = 0$ can  be found using the successive approximations method taking 6 roots $\pm x_j^{(0)}$ of the
equation $Y_r(x) = 0$ as a zeroth order approximation. In the first order approximation we add a small imaginary part to each root:
\begin{align}
x_j &= x_j^{(0)} +i\delta_j, \quad
 x_{j}^{(0)}=\pm x_{1}^{(0)},\ \pm x_2^{(0)},\ \pm x_3^{(0)}
\end{align}
After substituting this into Eq. (\ref{AligoY}) we find $\delta_j$:
\begin{align}
\notag
 Y_r'&\left(x_j^{(0)}\right)\,  2x_j^{(0)}\,i\delta_j +
 	ig x_j^{(0)}Y_i\left(x_j^{(0)}\right)=0,\\
\label{CondStab}
 \delta_j &= -\left(\frac{g}{2}\right)
  \frac{Y_i\left(x_j^{(0)}\right)}{Y_r'\left(x_j^{(0)}\right)}<00,
  \\
  \notag
 &\text{where }\
 	Y_r'\left(x_j^{(0)}\right)\equiv
 		\left.\frac {dY_r}{d (x^2)}\right|_{x^2=\left(x_j^{(0)}\right)^2}
\end{align}
In order to fulfill the condition for a stable oscillation,  $\delta_j$ have to be negative as we required in (\ref{CondStab})
(recall, as we assume the time dependence of position to be $\sim e^{-i\Omega t}$, hence, eigen frequency ($x_j$)
corresponds to oscillations $\sim e^{-i x_j t}$ and negative imaginary part means damping). Therefore, stable roots with $\delta_j < 0$
will be achieved in the case when the functions $Y_r'$ and $Y_i$ have the {\em same signs} in the regions close to
 roots of $Y_r$, i.e. their plots have to relate to each other as  shown on Fig.~\ref{stable}.

We would like to emphasize that the obtained criterion is very simple and that it allows us to estimate whether a certain set
of input parameters can provide roots (complex frequencies) with imaginary parts of the  desired signs without the need to
solve the characteristic equation. The proposed criterion is very close to the Raus-Gurwitz criteria \cite{68Korn}, however, it is more obvious
and more useful for the particular analysis of optical spring stability.

\begin{figure}[ht]
\psfrag{Yr}{$Y_r$} \psfrag{Yi}{$Y_i$} \psfrag{c}{\bf c)} \psfrag{d}{\bf d)}
\psfrag{x2}{$x^2$} \psfrag{x02}{$x_0^2$}
\psfrag{x01}{$\left[x_1^{(0)}\right]^2$}
\psfrag{x002}{$\left[x_2^{(0)}\right]^2$} \psfrag{x03}{$\left[x_3^{(0)}\right]^2$}
 \includegraphics[width=0.3\textwidth, height=0.15\textwidth]{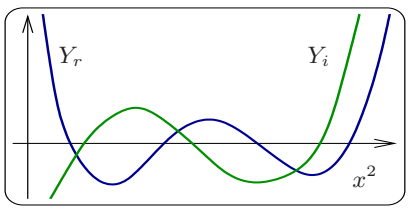}
 \caption{For the case of 3 pumps the roots are stable  if in accordance with criterion (\ref{CondStab}) the plots of $Y_r$ and $Y_i$ correspond to each other as shown.}\label{stable2}
\end{figure}

The stability criterion described in Eq. (\ref{CondStab}) can easily be generalized for a larger number of optical pumps.
For example, for three pumps the functions $Y_r$ and $Y_i$ are parabolas of fourth and third order correspondingly
(in respect to $x^2$) and the roots will be stable if the functions $Y_r'$ and $Y_i$ have the same sign at the regions close to the
 roots of $Y_r$  as shown in Fig.~\ref{stable2}.

Obviously, the regimes of stable optical spring considered in
\cite{08ChenPRD} fulfil the condition described in Eq. (\ref{CondStab}). However, our criterion might be useful to
identify configurations that can be experimentally implemented avoiding any feedback or for other interesting regimes
such as double resonance or  negative inertia \cite{10KhLIGO}.

\begin{table}[ht]
\caption{Parameters of Advanced LIGO used in this paper.}\label{table:LIGO}
\begin{tabular}{c|c}
Parameter & Value \\
\hline
Arm length, $L$ & $4$~km\\
Mass of each mirror, $m$ &$40$~kg\\
ITM amplitude transmittance, $T$ & $\sqrt {0.005}$\\
Relaxation rate of single FP cavity, $\gamma_0$ (\ref{gamma0})& $94$~s$^{-1}$\\
SRM amplitude reflectivity $\rho$& $\sqrt{0.93}$\\
Optical wavelength, $\lambda$ & $1064$~nm\\
\hline
\end{tabular}
\end{table}

\begin{figure}[t]
\psfrag{a}{\bf a)} \psfrag{b}{\bf b)}
\psfrag{Yr}{$Y_r$} \psfrag{Yi}{$Y_i$}  \psfrag{z3}{$x_3^2$}
\psfrag{barz1}{$\bar z_1$} \psfrag{barz2}{$\bar z_2$}
\psfrag{z0}{$x_0^2$} \psfrag{z}{$x^2$}
 \includegraphics[width=0.23\textwidth]{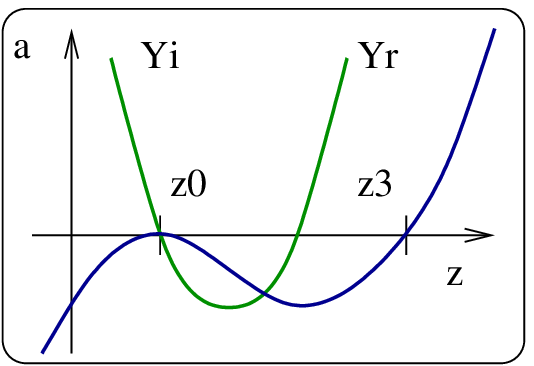}
 \includegraphics[width=0.23\textwidth]{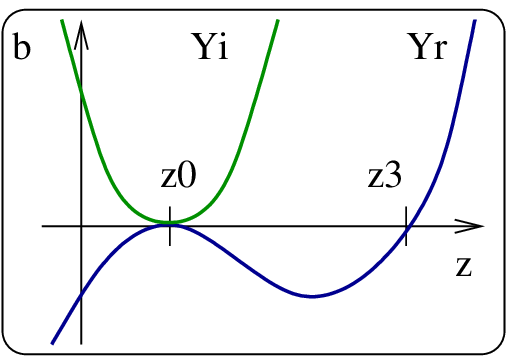}
 \caption{``Ideal'' double resonance: curve $Y_r$ toughes $z$-axis in point $z_0$: a) left root $\bar z_1$ of $Y_i$ coincides with $z_0$; b) curve $Y_i$ touches $z$-axis in the same point $z_0$.}\label{IdDR}
\end{figure}

\section{Stable Double Resonance}\label{SDR}

Oscillators in the double resonance  regime have interesting  properties.
The resonant force, with the  time dependence
$f=F_0\cos \Omega_\text{res}t$, acting on such an oscillator produces a displacement
$z_\text{dr}=-F_0t^2/(8m)\cos \Omega_\text{res}t\sim t^2$, which is much larger than the resonant
displacement  $z_\text{co}=-F_0t/(2m)\sin \Omega_\text{res}\sim t$ of a conventional oscillator under action
of the same force \cite{05a1LaVyPLA}. In the frequency domain the resonant gain of a conventional oscillator is
$\sim \Omega_\text{res}/\Delta \Omega$ (i.e. inverse proportional to the bandwidth $\Delta\Omega$ of
the resonance curve) whereas the resonant gain of an oscillator in the double resonance regime is much larger:
$\sim\big(\Omega_\text{res}/\Delta \Omega\big)^2$ (i.e. inverse proportional to the {\em square} of
the bandwidth $\big(\Delta\Omega\big)^2$). This feature, as mentioned in the introduction, is responsible
for the wide bandwidth gain in signal to noise ratio.

The double resonance takes place if the  susceptibility has a pole of second order \cite{01a1KhPLA,05a1LaVyPLA}.
 In terms of our stability criterion this means for the case of the double resonance that the plot of $Y_r$ as function
of $x^2$ touches the horizontal
axis at the resonance frequency $x_0^2$ and $Y_i$ is equal to zero at this point.  Fig.~\ref{IdDR} illustrates
two possible ways of behaviour of $Y_i$: in Fig.~\ref{IdDR}a and \ref{IdDR}b  $Y_i$ crosses and touches the horizontal axis at
$x^2=x_0^2$, respectively.
However, it is difficult to exactly realize  the  regimes shown in Fig.~\ref{IdDR} --- some discrepancies like shown
on Fig.~\ref{stable} are inevitable.

In general, double resonance takes place when any two of three roots $\big[x_1^{(0)}\big]^2,\ \big[x_2^{(0)}\big]^2,\
\big[x_3^{(0)}\big]^2$ coincide. In this section we consider the particular case of close smaller roots
 $\big[x_1^{(0)}\big]^2,\ \big[x_2^{(0)}\big]^2$ of $Y_r$
 (our preliminary analysis shows that the other case of close roots $\big[x_2^{(0)}\big]^2,\ \big[x_3^{(0)}\big]^2$ can not
provide a stable set of roots):
\begin{equation}
\label{closeRoots}
\left[x_{1,2}^{(0)}\right]^2\simeq  x_0^2\pm\epsilon\,, \quad \epsilon \ll 1
\end{equation}
where $x_0^2$ is a middle point between the roots $\left[x_{1,2}^{(0)}\right]^2$. Hence, $Y_r$ may be presented in the form
\begin{align}
\label{yrRoots}
 Y_r &= (x^2-x_0^2-\epsilon)(x^2-x_0^2+\epsilon)\left(x^2-\big[x_3^{(0)}\big]^2\right)
\end{align}

\begin{figure}[ht]
\psfrag{Yr}{$Y_r$} \psfrag{Yi}{$Y_i$} \psfrag{c}{\bf c)} \psfrag{b}{}
\psfrag{e}{$\epsilon$} \psfrag{e1}{$\epsilon_1$} \psfrag{e2}{$\epsilon_2$}
\psfrag{z}{$z=x^2$} \psfrag{z1}{$z_1$} \psfrag{z2}{$z_2$} \psfrag{z3}{$z_3$}
\psfrag{z0}{$z_0=x_0^2$}
 \includegraphics[width=0.35\textwidth, height=0.2\textwidth]{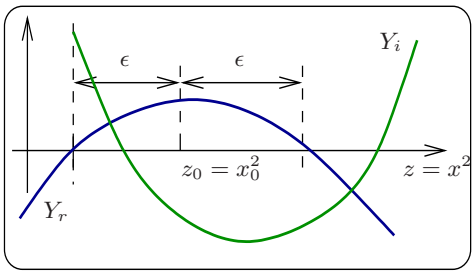}
 \caption{Detailed plot of two close roots of $Y_r$ (\ref{AligoYr}). The  roots of   $Y_i$  (\ref{AligoYi}) have to be shifted to the right relative to the roots of $Y_r$ in order to fulfill the stability criterion (\ref{CondStab}).} \label{2CloseRootsDet}
\end{figure}

In order to fulfill the criterion (\ref{CondStab}) one has to arrange the parabolas along the $x^2$ axis in
 such a way that the roots of $Y_i$ would be placed close to roots of $Y_r$ (see Fig.~\ref{2CloseRootsDet}).
Shifting $Y_i$ along the vertical axis allows us to manipulate the damping at the resonant frequencies, because the values of $Y_i$ at the roots of $Y_r$ define the mechanical damping of the oscillator (\ref{CondStab}). The problem of properly arranging this two curves  is solvable as there are enough parameters for manipulation available:
the two detunings and the two powers of the two pumps (i.e. the dimensionless parameters $P,\, Q,\,\eta,\, \nu$).
An example of a step by step calculation is presented in Appendix~\ref{DoubleRapp}. For the following dimensionless parameters
\begin{subequations}
 \label{DRparam}
\begin{align}
P&\simeq  0.77,\ Q \simeq -0.027,\quad g=0.6,\ d\simeq 1.39,\\
\eta& \simeq 0.24,\quad \nu=0.5,\quad \epsilon=0.05\,.
\end{align}
\end{subequations}
we get the eigen frequencies:
\begin{align}
 \Omega_1 &\simeq 532.3\, \text{s}^{-1},\quad \delta_1\simeq 16.8\ \text{s}^{-1},\\
 \Omega_2 &\simeq 482.1\, \text{s}^{-1},\quad \delta_2\simeq 15.4\ \text{s}^{-1},\\
 \Omega_3 &\simeq 961.8\, \text{s}^{-1},\quad \delta_3\simeq 314.2\ \text{s}^{-1},
\end{align}

\begin{figure}[ht]
\psfrag{P}{$|\Psi|$}
\psfrag{f}{$f$, Hz}  \psfrag{y}{$Psi(f)$}
\psfrag{g}[l][][.8]{$\epsilon$}
\psfrag{h}[l][][.8]{$\epsilon/2$}
\psfrag{t}[l][][.8]{$\epsilon/4$}
\psfrag{z}[l][][.8]{free mass}
 \includegraphics[width=0.45\textwidth, height=0.4\textwidth]{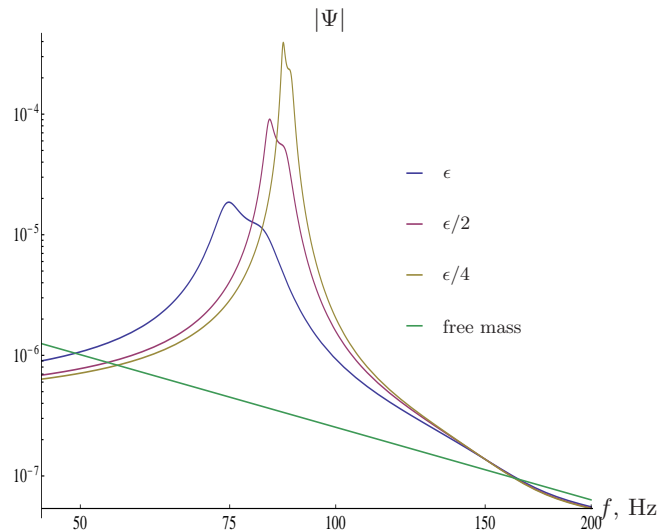}
 \caption{Trace 1: susceptibility $|\Psi(f)|$ as function of  frequency $f=\Omega/2\pi $
for the parameters given in Eqn (\ref{DRparam}). Traces 2,3: same as trace 1 but with %smaller
$\epsilon\to \epsilon/2$ and with $\epsilon \to\ \epsilon/4$.  Trace 4: susceptibility
 of free mass ($4/m(2\pi f)^2$). }\label{DpLigoPsi}
\end{figure}

Using Advanced LIGO parameters as listed in Table~\ref{table:LIGO}  the set (\ref{DRparam}) can be  recalculated, giving the following results:
\begin{align*}
\Gamma_1&\simeq 230\, \text{s}^{-1},\quad \Delta_1\simeq 1063\, \text{s}^{-1},\quad
 	I_1\simeq 863\, \text{kW},\\
 \Gamma_2 &\simeq 115\, \text{s}^{-1},\quad \Delta_2\simeq -526\, \text{s}^{-1},\quad
 	I_2\simeq 61\, \text{kW},\\
 &(\phi_1\simeq 1.49,\quad \phi_2\simeq -1.39),\
\end{align*}
Fig.~\ref{DpLigoPsi} shows the corresponding  susceptibility $\Psi^{-1}$ as a function of frequency
$f=\Omega/2\pi$ for the  listed parameters.
It is worth emphasising that it is possible to realize a stable double resonance
with {\em arbitrary small} damping by just choosing smaller values for $\epsilon$.
Fig.~\ref{DpLigoPsi} includes the susceptibility  for several values of parameter $\epsilon$: $\epsilon \to \epsilon/2,\
\epsilon/4$, as well as the susceptibility of the free masses as comparison.

\subsection{Possibilities of experimental observation}\label{PossEx}

It would be interesting to demonstrate and observe a stable double resonance in an experiment.
For example, this could be done in Gingin High Optical Power Test Facility \cite{Gingin06}.
The initial formulas (\ref{notations2}) are valid for the Gingin topology with one exception.
Only one arm of the Michelson interferometer (with Fabry-Perot arm cavities) is used in Gingin,
hence, one has to define the power parameters $P$ and $Q$ a factor $2$ smaller:
\begin{align}
 P & \equiv  \frac{16k_1I_1\Delta_1}{m L\big(\Gamma_1^2+\Delta_1^2\big)^2},\quad
 Q \equiv \frac{16 k_2I_2\Delta_2}{mL \big(\Gamma_1^2+\Delta_1^2\big)^2},
\end{align}
The Gingin facility features test masses of $m=0.8$~kg and arm length is $L=80$~m.
 Then, assuming $\Gamma_1=400\, \text{s}^{-1}$, it is easy to recalculate the parameters (\ref{DRparam}) for Gingin:
\begin{align}
 \Gamma_1&\simeq 400\, \text{s}^{-1},\quad \Delta_1\simeq 843\, \text{s}^{-1},\quad
 	I_1\simeq 3.59\, \text{kW},\\
 \Gamma_2 &\simeq 200\, \text{s}^{-1},\quad \Delta_2\simeq -911\, \text{s}^{-1},\
 	I_2\simeq 0.254\, \text{kW}.
\end{align}
It seems that these parameters may be relatively easy realized using the Gingin facility.

\begin{figure}[ht]
\psfrag{T1}{$T_\text{PRM}$}
\psfrag{T2}{$T_\text{IM}$}
\psfrag{PRM}{PRM}
\psfrag{ITM}{IM}
\psfrag{m}{$m$}
\psfrag{z}{$z$}
\psfrag{l}{$\ell$}
\psfrag{L}{$L$}
\includegraphics[width=0.45\textwidth]{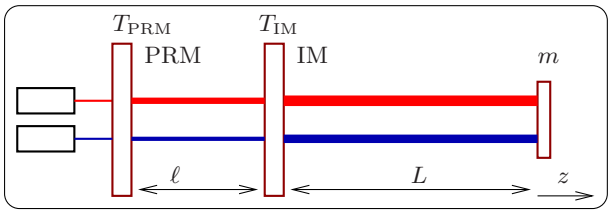}
\caption{Scheme of the optical rigidity experiment within the Glasgow Prototype Interferometer.
Only the end mirror can move, while the mirrors PRM and IM  can be considered as unmovable, because their masses are
 30 times larger than that of the end mirror.}\label{fig:GP}
\end{figure}

\begin{table}[ht]
\caption{Parameters of Glasgow Prototype Interferometer.}\label{table:GP}
\begin{tabular}{c|c}
Parameter & Value \\
\hline
Arm length, $L$ & $10$~m\\
Distance between PRM and ITM, $\ell$ & $5$~m\\
Mass of movable mirror, $m$ &$100$~g\\
Masses of PRM and IM, $M$ &$3$~kg\\
PRM amplitude transmittance $T_\text{PRM}$& $\sqrt{0.05}$\\
ITM amplitude transmittance, $T_\text{IM}$ & $\sqrt {0.01}$\\
Input power, $I_\text{input}$ & $\ge 1$~W,\\
Optical wavelength, $\lambda$ & $1064$~nm\\
\hline
\end{tabular}
\end{table}

Another experimental setup that can be used to observe a stable double resonance  is the Glasgow 10m Prototype \cite{Huttner07,Edgar09}.
In contrast to the balanced scheme of  a full Michelson interferometer such as Advanced LIGO, where it is possible to pump the
interferometer with the {\em symmetric} mode tuned to resonance, while observing the optical rigidity in the  {\em anti-symmetric} mode,
 in the Glasgow Prototype (shown on Fig.~\ref{fig:GP}) the same mode will have to be used to pump and to observe rigidity.
Recall that the smaller the detuning the larger is the circulating power, whereas optical rigidity requires a large detuning.
However, it is possible to set the first cavity between PRM and IM to anti-resonance
(the relaxation rates $\Gamma_1 = \Gamma_2=T_\text{PRM}T_\text{IM}c/4L$ reach their minimum) and to create the detuning by shifting
the position of the end mirror. In this situation the equations (\ref{notations2}) can be used with $\nu=1$. Taking into account
 that practically only the light end mirror can move the power parameters $P,\ Q$ have to be
 redefined, reduced by factor $4$:
\begin{subequations}
\label{PQGl}
\begin{align}
 P & \equiv  \frac{8k_1I_1\Delta_1}{m L\big(\Gamma_1^2+\Delta_1^2\big)^2},\quad
 Q \equiv \frac{8 k_2I_2\Delta_2}{mL \big(\Gamma_1^2+\Delta_1^2\big)^2}.
\end{align}
\end{subequations}
Then we can apply the scheme presented in Appendix~\ref{DoubleRapp}. In particular the stable double resonance
is characterised by the following set of dimensionless parameters:
\begin{subequations}
 \label{DRparamGL}
\begin{align}
P&\simeq  2.43,\ Q \simeq -0.51,\quad g=0.625,\ d\simeq 1.27,\\
\eta& \simeq 0.71,\quad \nu=1,\quad \epsilon=0.01\,.
\end{align}
\end{subequations}
Using  the actual parameters of the Glasgow Prototype, as presented in Table~\ref{table:GP},
we can calculate the following set of experimental parameters:
\begin{align}
 \Gamma_1&= \Gamma_2\simeq 937\, \text{s}^{-1},\quad
  \Delta_1  \simeq 4138\, \text{s}^{-1},\ \Delta_2  \simeq -3450\, \text{s}^{-1},\nonumber\\
 	&I_\text{1 input}  \simeq 1.28\, \text{W}, \quad I_\text{2 input}\simeq 0.33\, \text{W}.
\end{align}
Here $I_\text{1 input},\ I_\text{2 input}$ are the input powers of the two pump lasers.

Unfortunately, the observation of a stable double resonance  is  more difficult for smaller scale experiments. For example,
one may consider an experimental setup as described in \cite{07CorbitPRL} of a short Fabry-Perot cavity (without power recycling)
 and a light-weight movable
 mirror: the power transmittance of the input mirror is $T^2=0.8\times 10^{-3}$, the  distance between the  mirrors is $L\simeq 0.9$~m
and the mass of the movable mirror is $m=1$~g. Then the dimensionless parameters (\ref{DRparamGL}) can be recalculated as follows:
\begin{align}
 \Gamma_1&= \Gamma_2\simeq 66.7\times 10^3\, \text{s}^{-1},\\
  \Delta_1  &\simeq 294\times10^{3}\, \text{s}^{-1},\ \Delta_2  \simeq -245\times 10^3\, \text{s}^{-1},\\
 	&I_\text{1 input}  \simeq 2.64\, \text{kW}, \quad I_\text{2 input}\simeq 1.26\, \text{kW}.
\end{align}
Obviously these input power values are impermissibly huge to be realised in an experiment. In order to find some
regularities concerning the experiment scale we can rewrite the formula \eqref{PQGl} in the  following form using the value of input power:
\begin{equation}
\label{Pest}
P = \left( \frac{8 k_1 L^2 I_\text{1 input}}{m c^3 (T_\text{eff}/4)^4} \right) \times \frac{ \Delta_1 / \Gamma_1}{(1 + \Delta_1^2/\Gamma_1^2)^3} \propto \frac{L^2 I_\text{1 input}}{m T_\text{eff}^4},
\end{equation}
where we substitute $\Gamma_1 = T^2_\text{eff}c /4 L$. For the  Glasgow prototype  the effective transmittance is
equal to $T_\text{eff}=T_\text{PRM}T_\text{IM}/4$.

In order to find out whether an experiment is realizable, we emphasized in expression (\ref{Pest}) only
the
proportionality to the terms, which  significantly change with changes of the experiment scale. Decreasing the scale
of the experiment is usually represented by decreasing the mass $m$ of the movable mirror and the distance $L$ between
the mirrors, which approximately compensate each others contribution into $P$. That means that in order to achieve
the  needed value of $P$ and thus to observe  the desired stable double resonance one needs to keep the effective transmittance
roughly the same as in a bigger scale experiment due to the strong dependence of $P$ on $T_\text{eff}$. This leads to the
need for the circulating power to be the same as for the bigger scale experiment, which would be pretty difficult due to problems
 caused by absorption induced heating of the small mirror.

\section{Conclusion and Outlook}
We developed a new criterion for the stability of an optical spring and applied it to the double resonance scheme. Before such a concept can be realised in a large scale gravitational wave detector it would be beneficial to demonstrate it on a prototype scale experiment, such as the Glasgow Prototype
\cite{Huttner07,Edgar09}, Gingin High Optical Power Test Facility \cite{Gingin06} or the AEI 10m Prototype Interferometer \cite{AEI10}.

It is worth noting that our criterion for optical spring stability may  also be applied to the negative inertia regime recently proposed by F.Ya.~Khalili and colleagues \cite{10KhLIGO} (the term ``negative inertia'' was proposed earlier by H.~M\"uller-Ebhardt \cite{08Muller}). This directly follows from the fact that the   negative inertia scheme formally corresponds to a double resonance configuration with a resonance frequency of zero.

The formulated stability criterion is also valid for gravitational wave detectors pumped by more than two lasers.
This may allow us to realize {\em stable triple (or quadro)} resonance, i.e. regimes when three (or four) eigen-frequencies
 coincide or are close to each other. In this case the response to force at the resonance frequency
is greater than for a double resonance configuration.  In the frequency domain the resonant gain for a triple resonance has
 to be proportional to $\sim\big(\Omega_\text{res}/\Delta \Omega\big)^3$ (i.e. inverse proportional to the  {\em third} power
 of the  bandwidth $\big(\Delta\Omega\big)^3$). It will allow us to further increase the gain in the  signal to noise ratio for wide
bandwidth signals, mentioned in the Introduction, by a factor $\Omega_\text{res}/\Delta \Omega$ compared to the double resonance. In this paper no sensitivity analysis for gravitational detectors was performed, but we plan to do this in a future publication.

An stable double resonance featuring arbitrarily large susceptibility may allow us to use it for the production of squeezed light
by means of the ponderomotive nonlinearity
\cite{06CorbitPRA,07CorbitPRL}. However, the obtained squeezing will have specific features \cite{08ChenPRD}: One would have to measure
two optimal quadratures, corresponding to each carrier, and squeezing can only be observed by finding  their correct combination, because
 they are correlated (entangled)  through the mechanical degree of freedom.

Due to the  combination of a uniquely large susceptibility and a low noise level the stable double (or triple) resonance
may also be applied to macroscopic quantum mechanics experiments with mirrors of relatively small mass. In particular, it can be used
for the observation of quantum entanglement between an oscillator
and an e.m. field  \cite{10a1MiDaChPRA,10a1MiDaMuReSoChPRA} or for preparation of the mechanical oscillator in a non-gaussian quantum state
\cite{10a1KhDaMiMuYaChPRL}.

In addition stable double resonance may be a useful instrument in other precision measurements, such as the  measurements of thermal
noise of mirror coatings.
 In case the mass $m$ of one cavity mirror is small enough compared to the other one,   the  equation for the variation
 $z$ of the optical distance inside the cavity
may be written in the frequency domain as follows:
\begin{align}
 m\big(-\Omega^2+K(\Omega)\big)\, z &= F_\text{ba} - m\Omega^2\,z_\text{th}\, ,
\end{align}
where $z_\text{th}$ are the thermal fluctuations of the  mirror surface in respect to the mirror's center of mass
 and $F_\text{ba}$ is the back action
force. Homodyne detection of the  output wave yields the  information on $z$. One can obtain resonant gain of the signal
 (in this case it is $z_\text{th}$) inside the resonance bandwidth
(i.e. for $\Omega=\Omega_\text{res}\pm \Delta\Omega/2$) and in case of a stable double resonance this gain is unusually large.

\acknowledgments
The authors are grateful for fruitful discussion with Farid Khalili, Stefan Danilishin and Kenneth Strain.
A.R's and S.V.'s research has been supported  by the Russian Foundation for Basic Research Grant No.
08-02-00580-a and NSF grant PHY-0651036. S.H. is supported by the Science and Technology Facilities Council
(STFC).

\appendix

\section{Details of the double resonance stability analysis}\label{DoubleRapp}

In this appendix we present the example estimation of a set of input parameter values needed to realize
the regime of a stable double resonance. We will deal mainly with the functions $Y_r$ and $Y_i$
in this section. As it was mentioned in Sec.~\ref{SDR} we need
to arrange these parabolas along the $x$-axis to achieve close stable frequencies and then arrange the parabola $Y_i$ along
the
$y$-axis to obtain the desired values of the imaginary parts of the frequencies or, other words, the desired bandwidth of
the double resonance.

First we assume the two smaller roots of $Y_r$ to be close enough to each other. That means that the function $Y_r$ can be expressed in
form \eqref{yrRoots}
\begin{equation}
\label{yrRootsL}
 Y_r= \big(z-z_0-\epsilon\big) \big(z-z_0+\epsilon\big)\big(z-z_3\big), \
 	z\equiv x^2,
\end{equation}
where $ 2 \epsilon$ is the distance between the close roots (and $ \epsilon$ is small). We also introduce the parameter $d = z_3 - z_0$
which is defined as the distance between the center of the two close roots and the ``lonely'' root.

Proper substitution transforms this cubic function into the reduced form $Y_r = y^3 + p y + q$. In our case the substitution
is $z = y + ( 2z_0+z_3)/3$. The particular case of this equation with $z = z_0$ and $y = y_0$ ($y_0$ is the middle of the close roots
in terms of the shifted variable) leads us to the following equation:
\begin{equation}
\label{eq.y0d}
y_0 = - \frac{z_3 - z_0}{3} = - \frac d 3.
\end{equation}

After the substitution is done the parameters $p$ and $q$ are expressed in terms of $\epsilon$ and $d$  as
follows:
\begin{equation}
\label{pqz1z2L}
 p = - \frac{1}{3}\, d^2 -\epsilon^2, \quad q = - \frac{2}{27}d \big(d^2-9\epsilon^2\big).
\end{equation}

On the other hand the function $Y_r$ is expressed \eqref{AligoYr} in terms of the input parameters such as the pump powers and the detunings
 (parameters $P, \ Q, \ g, \ \eta, \ \nu$). Substitution of
\begin{equation}
\notag
y=z + \frac{ 2 (1 + \eta ) + \nu g^2}{3}
\end{equation}
transforms the function into  the familiar form $Y_r = y^3 + p y + q$. However, this time the coefficients $p$ and $q$ are expressed in terms
of the above mentioned input parameters.

Equating the pairs of expressions for $p$ and $q$ (expressed in different terms) one can obtain the solutions for the pump powers $P$ and $Q$
as functions of the remaining input parameters as well as $\epsilon$ and $d$.

At this step the $x$-arrangement of the function $Y_r$ is completed. Indeed, if we set a certain value of $d$ we can
expect the middle of the close roots
(in the shifted variables) to be equal to $-d/3$. Setting a certain value of $\epsilon$ results in the close roots to be separated by  $2 \epsilon$.

\begin{figure}[ht]
\psfrag{Yr}{$Y_r$} \psfrag{Yi}{$Y_i$} \psfrag{c}{\bf c)} \psfrag{b}{}
\psfrag{e}{$\epsilon$} \psfrag{e1}{$\epsilon_1$}
\psfrag{z}{$z$} \psfrag{z1}{$z_1$} \psfrag{z2}{$z_2$} \psfrag{z3}{$z_3$}
\psfrag{z0}{$z_0$}
 \includegraphics[width=0.33\textwidth, height=0.25	\textwidth]{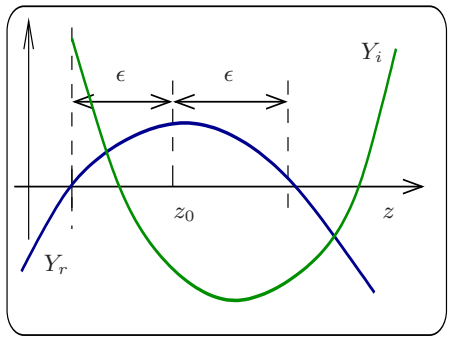}
 \caption{Detailed plot of the two close roots of $Y_r$ (\ref{AligoYr}): $z_{1,2}=z_0\pm
\epsilon$. The  roots of   $Y_i$  (\ref{AligoYi}) are separated by the same distance $2\epsilon$,
 but the trace  of $Y_i$ itself is shifted by the distance $\epsilon_1$.}
\label{TriRCloseLigoDel}
\end{figure}

The next step is to  arrange  $Y_i$ thus it provides stability of the complex frequencies.
According to the stability criterion~\eqref{CondStab} frequencies are expected to be stable if the
roots of $Y_r$ ($z_1, \ z_2$) and $Y_i$ ($\bar z_1,\ \bar z_2$) are located in the same way as is shown on Fig. \ref{TriRCloseLigoDel}:
\begin{equation}
\label{eq.zzbar}
   z_1 < \bar z_1,\quad z_2<\bar z_2.
\end{equation}
The values $z_{1,2}$ can be estimated using Cardano's formulae (see Appendix \ref{Cardano}) and $\bar z_{1,2}$ are
the roots of a quadratic equation:
\begin{align}
\notag
   z_{1,2} & \simeq -\left(- \frac{q}{2} \right)^{1/3}+ \frac{2 (1+\eta )+\nu g^2}{3} \mp \epsilon,
   \\
   \notag
  \bar z_{1,2} &= \frac{ \nu + \eta }{ 1 + \nu } \mp
  	\sqrt{ \left( \frac{ \nu + \eta }{ 1 + \nu } \right)^2 - \frac{\nu P + Q }{ 1 + \nu}}\, .
 \end{align}
After properly  arranging $Y_i$ thus it complies with the condition \eqref{eq.zzbar} the second step is completed.
We obtained three stable frequencies, two of which are separated by the desired distance of $2 \epsilon$.

It is obvious that in order to obtain a single resonance peak, one should choose a distance between the close frequencies smaller than the mean
 imaginary part of these roots $\delta_j$. Since $z = x^2$, the distance between the close frequencies can be estimated as
 $x_2 - x_1 \sim \epsilon / \sqrt{z_0}$. Equantion \eqref{CondStab} gives an estimate for $\delta$.
Summing this up one can write
\begin{equation}
\notag
 \frac{\epsilon}{\sqrt{z_0}} \lesssim \left| \frac{ g Y_i (z_{1,2}) }{2Y_r'(z_{1,2})}\right|\simeq 	\left| \frac{ g Y_i (z_{1,2}) }{4 \epsilon d}\right|,
\end{equation}
or
\begin{equation}
 \label{DiffRoots2}
 \big|Y_i(z_{1,2}) \big|  \ge \frac{4\epsilon^2 d}{g_1\sqrt{z_0}}\,.
\end{equation}

For us regimes with infinitely close frequencies, i.e.  with $\epsilon \rightarrow 0$, are of interest.
 Along with decreasing the distance between
frequencies we can also decrease the bandwidth of the double resonance or in other words the values of $\delta_j$, which is possible by decreasing the values of $Y_i$. Equation \eqref{DiffRoots2} is the only condition these values should obey.

To ensure this we apply the following conditions to the values of $Y_i$:
\begin{equation}
\label{CondYi}
 Y_i(z_1)  =  s_1\,\frac{4\epsilon^2 d}{g_1\sqrt{z_0}}, \quad Y_i(z_2)  = - s_2\,\frac{4\epsilon^2 d}{g_1\sqrt{z_0}},
\end{equation}
where the variables $s_1$ and $s_2$  are subject to fitting and should be approximately equal to unity.

This is the final step, resulting in a set of three stable frequencies, two of which form the resonant peak of the desired bandwidth
in the susceptibility.

It is worth noting that the  parameter $\epsilon$ is proportional to distance the between the roots of $Y_r$ and, hence, $\epsilon$ is proportional
to the  bandwidth of the resonance. The parameters $g,\ \nu$ are free and can be used to vary the resonance frequencies or the
values of the pumping powers.

\section{Cubic equation: Cardano's formula}\label{Cardano}

The cubic equation
\begin{equation}
 x^3+ax^2+bx+c=0
\end{equation}
may be written in the depressed cubic form by substitution of $x=y-a/3$ \cite{68Korn}:
\begin{align}
 0&= y^3 +py +q,\\
 p& =-\frac{a^2}{3} +b,\quad
 q= 2\left(\frac{a}{3}\right)^3 -\frac{ab}{3} +c
\end{align}
The roots of this depressed cubic equation are given by:
\begin{subequations}
\label{CardanoSol}
\begin{align}
 y_{1,2}&= -\frac{A+B}{2}\pm  i\sqrt 3\, \frac{A-B}{2}, \quad y_3= A+B, \\
 &\text{where}\nonumber\\
 A &= \left(-\frac{q}{2} +\sqrt{D}\right)^{1/3},\quad
 B=\left(-\frac{q}{2} -\sqrt{D}\right)^{1/3},\\
 D&=\left(\frac{p}{3}\right)^3 + \left(\frac{q}{2}\right)^2
\end{align}
\end{subequations}
For $A$ and $B$ one should choose any cubic roots from the corresponding expressions, which
fulfill the following equation:
\begin{equation}
 AB=-p/3.
\end{equation}
In case this equation has real coefficients  one should take the real values of the roots (if possible).

%\bibliography{Rigidity}
\end{document}